\documentclass[runningheads]{llncs}
\usepackage[T1]{fontenc}

\usepackage{amsmath}
\usepackage{float}
\usepackage{times}
\usepackage{url}

\usepackage{breakurl}
\usepackage[breaklinks]{hyperref}
\usepackage{booktabs}
\usepackage{todonotes}
\usepackage{parskip}

\usepackage{listings}

\usepackage{graphicx}
\usepackage{caption}
\usepackage{subcaption}
\begin{document}
\title{Understanding the Energy Consumption of HPC Scale Artificial Intelligence}
%
%
\author{Danilo Carastan-Santos\inst{1}\orcidID{0000-0002-1878-8137} \and Thi Hoang Thi Pham\inst{1}}
\authorrunning{Carastan-Santos and Pham}
%
\institute{Univ. Grenoble Alpes, CNRS, Inria, Grenoble INP, LIG\\
Grenoble, France\\
\email{danilo.carastan-dos-santos@inria.fr, hoangthi.phamthi@gmail.com}}
\maketitle              
\begin{abstract}

This paper contributes towards better understanding the energy consumption trade-offs of HPC scale Artificial Intelligence (AI), and more specifically Deep Learning (DL) algorithms.
For this task we developed benchmark-tracker, a benchmark tool to evaluate the speed and energy consumption of DL algorithms in HPC environments. 
We exploited hardware counters and Python libraries to collect energy information through software, which enabled us to instrument a known AI benchmark tool, and to evaluate the energy consumption of numerous DL algorithms and models.
Through an experimental campaign, we show a case example of the potential of benchmark-tracker to measure the computing speed and the energy consumption for training and inference DL algorithms, and also the potential of Benchmark-Tracker to help better understanding the energy behavior of DL algorithms in HPC platforms. 
This work is a step forward to better understand the energy consumption of Deep Learning in HPC, and it also contributes with a new tool to help HPC DL developers to better balance the HPC infrastructure in terms of speed and energy consumption.

\end{abstract}
\section{Introduction}

The current direction in Artificial Intelligence, and more specifically Deep Learning (DL), is clearly to orders of magnitude more compute~\cite{aigrowth}, reaching High-Performance Computing (HPC) scale. That means more energy, which comes from sources such as fossil fuels, nuclear power, water dams, wind, \textit{etc.}. Fossil fuel is the main source, contributing to 36\%\footnote{\url{https://ec.europa.eu/eurostat/cache/infographs/energy/bloc-3b.html?lang=en}} in the total energy sources mix. Fossil energy emits a significant amount of CO2 into the environment. Under all of these observations, it is therefore important to monitor the energy consumption of DL, to master its energy demand and attenuate its contribution to climate change.


Typical DL research focuses mainly in the quality of the predictions of a trained DL model. 
As Deep Learning being a significant part of AI, understanding DL and its energy consumption will build us the path to better balance computing and energy resources needed for its proper operation, and thus being less energy demanding. 

This paper is a step towards building this path. It addresses the following challenges:

\begin{itemize}
    \item \textit{For Deep Learning running in HPC platforms, how much energy are the current popular and widely used DNNs consume? }
    \item \textit{Is it accurate to say that: More complex models will cost more energy?}
    \item \textit{Does the model give higher accuracy, more energy will be consumed?}
\end{itemize}
In this work, we instrumented a Deep Learning benchmark with a software energy measurement tool to output the Benchmark-Tracker, which tracks the energy consumption of DNN models insight the DL benchmark. The results from running experiments with the developed instrument give us a better understanding of today's energy consumption for the widely used DNN models. From those insights, we can expand to further and more in-depth future studies on energy consumption issues.
The available version of Benchmark-Tracker is on GitHub \footnote{\url{https://github.com/phamthi1812/Benchmark-Tracker}}.

We organized the remaining of this paper in the following manner: Section~\ref{sec:related} presents the related works. In Section~\ref{sec:background} we present some preliminary background information. Section~\ref{sec:methods} briefly presents the instrumentation details to implement Benchmark-Tracker, and Section~\ref{sec:results} present some preliminary results of our tool. Finally, in Section~\ref{sec:conclusion} we present our concluding remarks and our planned future works.

\section{Related work}
\label{sec:related}

\subsection{AI and climate change}
Climate change is a crucial issue for people all over the world. According to experts in the field, AI has the potential to accelerate the process of environmental degradation. For example, large-scale natural language processing models -- specifically transformer~\cite{vaswani2017attention} models -- have a huge carbon footprint \cite{hurting}. Alternatively, ``There is a real need to think about how you're building these systems. Are you training a needlessly complex algorithm? How frequently are you retraining?''\cite{hurting}. In order to understand energy consumption grounds, firstly, we have entered the background detailed in \cite{eximpacttracker}. Secondly, we have to understand that not only using electrical energy to train an DNN creates CO2, but also collecting and storing data. As DL becomes more complex, data centers are essential for storing large amounts of data needed to power DL systems, but require significant energy. ``Data centers are going to be one of the most impactful things on the environment''\cite{gpu}. Additionally, training advanced artificial intelligence systems, including deep learning models, may require high-powered GPUs running for days at a time and GPUs that use much power to run machine learning training have contributed to significant CO2 emissions.\cite{gpu_ex}.

\subsection{Energy-Aware AI}
Many practical aspects were attempted by numerous works \cite{Mazouz2017AnIM,GarcaMartn2019EstimationOE,codecarbon,ficher:hal-03196527}. Besides focusing on reducing energy consumption in the training process, one of the possibilities is that we can have better energy management once its energy consumption can be modeled and predicted. Chen \textit{et al.}~\cite{en_using_deep} proposed the idea of using deep learning to model energy consumption. In technical points, the hardware provider has also tried to give the better compatible hardware or especially GPU, which gives the best performance in \cite{effortgpu}.

As some ideas have been outlined above, there are numerous efforts to better control energy consumption, but they still do not reduce the development and power of Deep Learning. We can consider when training models, such as controlling the number of parameters in the model or finding a better efficient way to store data which has mentioned in \cite{gooddata} or \cite{mixedimapct}. 

\subsection{AI Benchmarks}
There are many benchmarks available today for AI, for example, AI Benchmark\footnote{\url{https://ai-benchmark.com/alpha}} works for both mobile devices and desktops, Dynabench\footnote{\url{https://dynabench.org/}} mainly works on Natural language, Sentiment Analysis, or Cloud-Based AutoML\footnote{\url{https://cloud.google.com/automl/}}, which can enable developers with limited machine learning expertise to train high-quality models specific to their needs. Each tool is developed with an emphasis on a fixed factor, notably the quality of the models' predictions.

\subsection{Energy Measurement Tools}
There are several tools based on three methodology points: Estimation from hardware characteristics (Green algorithms, ML CO2 Impact), External measures from outside the hardware (Wattmeters), or Software power models from hardware performance counters (CodeCarbon, Experiment-Impact-Tracker \cite{eximpacttracker}, CarbonTracker, Energy Scope) \cite{tools}.

\subsection{Positioning of this paper}
Our paper situates in two fronts: (i) Our paper bridges AI benchmarks and energy measurement tools, giving an out-of-the-box tool to help HPC DL developers to better balance the HPC infrastructure in terms of speed and energy consumption, and (ii) we go beyond only evaluating the prediction quality of AI models and algorithms, but we also evaluate the energy/complexity/performance trade-offs of popular AI models and algorithms, taking into consideration several HPC hardware.


\section{Background}
\label{sec:background}

In this section we present information about the used tools. We start by bringing details about the AI Benchmark Alpha. We then bring some details about measuring/estimating the energy consumption and CO2 emissions of HPC platforms, and we finish by presenting details about the Experiment-Impact-Tracker tool.


\subsubsection{AI Benchmark Alpha}
The AI Benchmark Alpha is an open-source\footnote{\url{https://pypi.org/project/ai-benchmark/}} library, for evaluating the AI performance of various hardware platforms, including CPUs, GPUs, and TPUs. The benchmark relies on the TensorFlow~\cite{tensorflow2015-whitepaper} machine learning library and provides a precise and lightweight solution for assessing inference and training speed for widely used and popular Deep Learning models. 

AI Benchmark treats the training and inference of the models as \textit{tests}. The tests cover all major Deep Learning models and algorithms. They include: Classification (MobileNet-V2, Inception-V3, Inception-V4, Inception-ResNet-V2, ResNet-V2-50, ResNet-V2-152, VGG-16, Image-to-Image Mapping (SRCNN 9-5-5, VGG-19, ResNet-SRGAN, ResNet-DPED, U-Net, Nvidia-SPADE), Image Segmentation (ICNet, PSPNet, DeepLab), Image Inpainting (Pixel-RNN), Sentence Sentiment Analysis (LSTM), and Text Translation (GNMT). 


\subsubsection*{Energy consumption and carbon emissions in HPC platforms}

When we measure the energy consumption for training an AI model running in HPC platforms, we also have to consider the additional power used to run the platform that is not directly related to computing, such as cooling. We call this power overhead as Power Usage Effectiveness (PUE), which is a number that depends on the HPC platform cooling efficiency and is often slightly larger than 1, and acts as a multiplicative factor for the measured energy consumption of the computing nodes.


From the energy level consumed obtained at the computing nodes, and multiplied by the PUE, we can calculate the corresponding CO2 emissions released into the environment as follows:
\begin{equation*}
    Emission_{carbon} = Energy_{computing} \times Intensity_{carbon}
\end{equation*}
The intensity of carbon is the number of grams of carbon dioxide (CO2) that it takes to make one unit of electricity a kilowatt per hour (kWh). 

The quantity of carbon emission is just as substantial as the energy consumption because the higher the carbon intensity, the more polluting the energy consumption. The carbon intensity of electricity generation depends on the energy sources mix, and it varies from region to region\footnote{\url{https://app.electricitymap.org/map}}. For our experiments we adopted 55g CO2/kWh, which relates to France's typical carbon intensity\footnote{\url{https://app.electricitymap.org/map}}.



\subsubsection{ The Experiment-Impact-Tracker}
The Experiment-Impact-Tracker is a software tool\footnote{\url{https://github.com/Breakend/experiment-impact-tracker}} that provides a simple plug-and-play solution for tracking your system's energy use, carbon emissions, and compute utilization. It records: power consumption from CPU and GPU, hardware information, and projected carbon emission information on Linux computers with Intel CPUs that implement the Running Average Power Limit (RAPL) and NVIDIA GPUs. 

Through the following example, we see how one can track the energy consumption by covering the process with the Experiment-Impact-Tracker:

\begin{small}
\begin{lstlisting}[language=Python]
from experiment_impact_tracker.compute_tracker import ImpactTracker
from experiment_impact_tracker.data_interface import DataInterface
os.mkdir("give_a_path")
tracker = ImpactTracker("given_path")
tracker.launch_impact_monitor()
Put_Your_Process_here()
tracker_results = {}
data_interface = DataInterface(["given_path"])
\end{lstlisting}
\end{small}

 The Experiment-Impact-Tracker launches a separate python process that gathers the energy consumption information in the background. Also, we can then access the information via the DataInterface. Moreover, like we mentioned before about how useful this Tracker is because of this context management, as illustrated in the listing below:
\begin{lstlisting}{python}
experiment1 = tempfile.mkdtemp()
experiment2 = tempfile.mkdtemp()

with ImpactTracker(experiment1):
    do_something()

with ImpactTracker(experiment2):
    do_something_else()
\end{lstlisting}

The combination of the AI Benchmark Alpha with the Experiment-Impact-Tracker that we did to produce Benchmark Tracker, which is used for running the experiments to understand the energy behavior of AI algorithms, is presented in the following subsection.

\section{Benchmark Tracker}
\label{sec:methods}


The essence of Benchmark-Tracker is in the instrumentation of the AI Benchmark Alpha with Experiment-Impact-Tracer, to also measure the energy consumption.
For this task, we identified the code regions where AI Benchmark runs for each model's training and inference. From there, we instrumented the Experiment-Impact-Tracker in these code regions, intending to be able to measure not only the hardware benchmark from the AIBenchmark but also the energy consumption of each model running in this benchmark. We can control which process we would like to run in the primary run file. The process goes as follows:
\begin{enumerate}
    \item Benchmark runs and starts calling the corresponding tests for each model.
    \item The Experiment-Impact-Tracker is activated and starts measuring energy during the training process of that model.
    \item The Tracker is turned off, and data logging occurs.
    \item The inference process of the same model happens (if we run this tool for both training and inference).
    \item Another test is called up to continue the process.
\end{enumerate}

The inference process is shortly presented as follows: 

\begin{lstlisting}[language=Python]
for subTest in (test.inference):
    os.mkdir(PATH)                 
    with ImpactTracker(PATH):
       <<TRACKED_CODE>>
    tracker_INFERENCE_results = {}
    data_interface = DataInterface([PATH])
\end{lstlisting}

The advantage of using experiment-impact-tracker becomes evident, since we can easily instrument a code section using the Python context management (i.e., the \texttt{with} statement). In the case above, \texttt{<<TRACKED\_CODE>>} refers to the code commands for an inference task. The same instrumentation procedure holds for the training tasks.

With the AI benchmark's dataset and our evaluated hardware (see Section 5.1), the AI benchmark runs approximately 60 seconds for one model. The Experiment-Impact-Tracker provides inaccurate estimations when the processing time is too short (in the order of 60 seconds).
With this in mind, we artificially increase the size of the dataset to increase the processing time, and have accurate energy measurements.
Artificially increasing the size of the dataset invalidate the accurate computed at the training tasks. That is why in the next section we compare the model's accuracy in terms of their reported accuracy in ImageNet\footnote{\url{https://ai-benchmark.com/tests.html}}. We plan to release a new version of Benchmark-Tracker with an appropriate dataset in future work (see Section~\ref{sec:conclusion}).

\section{Results}
\label{sec:results}
\subsection{Experimental setting}
For the experiments, we used Grid'5000\footnote{\url{https://www.grid5000.fr/w/Grid5000:Home}} High-Performance Computing test-bed. 
We used the Chifflet node (Model: Dell PowerEdge R730, CPU: Intel Xeon E5-2680 v4, Memory: 768 GiB, GPU: 2 x Nvidia GeForce GTX 1080 Ti (11 GiB)).

\subsection{Experimental results}
We ran Benchmark-Tracker 10 times to achieve the below statistical results.
The bar plots below represent an estimate of the central tendency for energy consumption with the height of each rectangle. It provides some indication of the uncertainty around that estimate using error bars. Intuitively, we see that the confidence intervals bar graph shows a slight error between 10 sets of the outcomes. It is also essential to remember that a bar plot shows only the mean value.

Let recall the definition of Training and Inference. We can say Training and Inference are the norm in DL. They are  two key processes associated with developing and using AI:
\begin{itemize}
    \item Training is ``teaching" a  Deep Neural Network (DNN) to perform the desired AI task (such as image classification or converting speech into text). We can express that it fits a model for training data. During the DL training process, the data scientist is trying to guide the DNN model to converge and gain the expected accuracy.
    \item The inference uses a trained DNN model to make predictions against previously unseen data and perform decision-making. The DL training process involves inference, because each time an image is fed into the DNN during training, the DNN tries to classify it. Given this, people usually deploy a trained DNN for inference. For example, one could make a copy of a trained DNN and start using it “as is” for inference. 
\end{itemize}
Therefore, the results are grouped into training and inference phases. 

With AI tasks, created DNN models can be large and complex, with dozens or hundreds of layers of artificial neurons and millions or billions of weights linking them. Normally, the bigger the DNN, the more computing, memory, and energy are consumed to execute it, and the longer will be the response time (or “latency”) from when you input data to the DNN until you obtain an outcome. 

Without losing generality, to better focus the analysis we only access the results of the classification task. It is important to remind that, thanks to AI Benchmark Alpha, Behcnkmark-Tracker also evaluates other kinds of tasks, such as (Classification, Image-to-Image Mapping, Image Segmentation). Table~\ref{tab:complex} presents the classification models' complexity, measured in the number of parameters.
\begin{table}[!h]
\caption{Image Classification Model Complexity, measured as the number of parameters \cite{Bianco_2018} }
\label{tab:complex}
\centering
\begin{tabular}{l|c}
  \toprule 
  \textbf{Model} & \textbf{Number of parameter (M: millions)}\\
  \midrule 
    MobileNet-V2 & 5M\\
    Inception-V3 & 25M\\
    Inception-V4 & 35M\\
    Inception-ResNet-V2 & 60M\\
    ResNet-V2-50 & 30M\\
    ResNet-V2-152 & 70M\\
    VGG-16 & 150M\\
  \bottomrule 
\end{tabular}
\end{table}

Also, to give the reader an impression of the connection between the structure of the DNNs and the energy consumption, we briefly present the main ideas of the evaluated models for the image classification task:
\begin{itemize}
    \item MobileNet-V2: Depth wise Separable Convolution which dramatically reduces the complexity cost and model size of the network.
    \item Inception-V3: Factorizing Convolutions, which reduces the number of connections/parameters without decreasing network efficiency, helps realize computational efficiency and fewer parameters. 
    \item Inception-V4: A more uniform, simplified architecture and more inception modules than Inception-v3. It uses asymmetric filters.
    \item Inception-ResNet-V2: Use a part of Inception-V4 and replace connection by residual links.
    \item ResNet-V2-50: It introduces skip connection (or shortcut connection) to fit the input from the previous layer to the next layer without any modification of the input. This version has 50 layers and uses residual links.
    \item ResNet-V2-152: Same idea with ResNet-V2-50, but this one has the maximum number of layers 152. ResNet is the Winner of ILSVRC 2015 in image classification, detection, and localization, as well as Winner of MS COCO 2015 detection, and segmentation.
    \item VGG-16: by using $3 \times 3$ filters uniformly, VGG-16 reduces the number of weight parameters in the model significantly. It helps in reducing the complexity of computing.
\end{itemize}
We describe the results in inference belonging to each type of model. Additionally, the rectangle's color shows the reported accuracy on ImageNet for each of the evaluated models\footnote{\url{https://ai-benchmark.com/tests.html}}.

\begin{figure}[!h]
	\centering
	\includegraphics[width=1\linewidth]{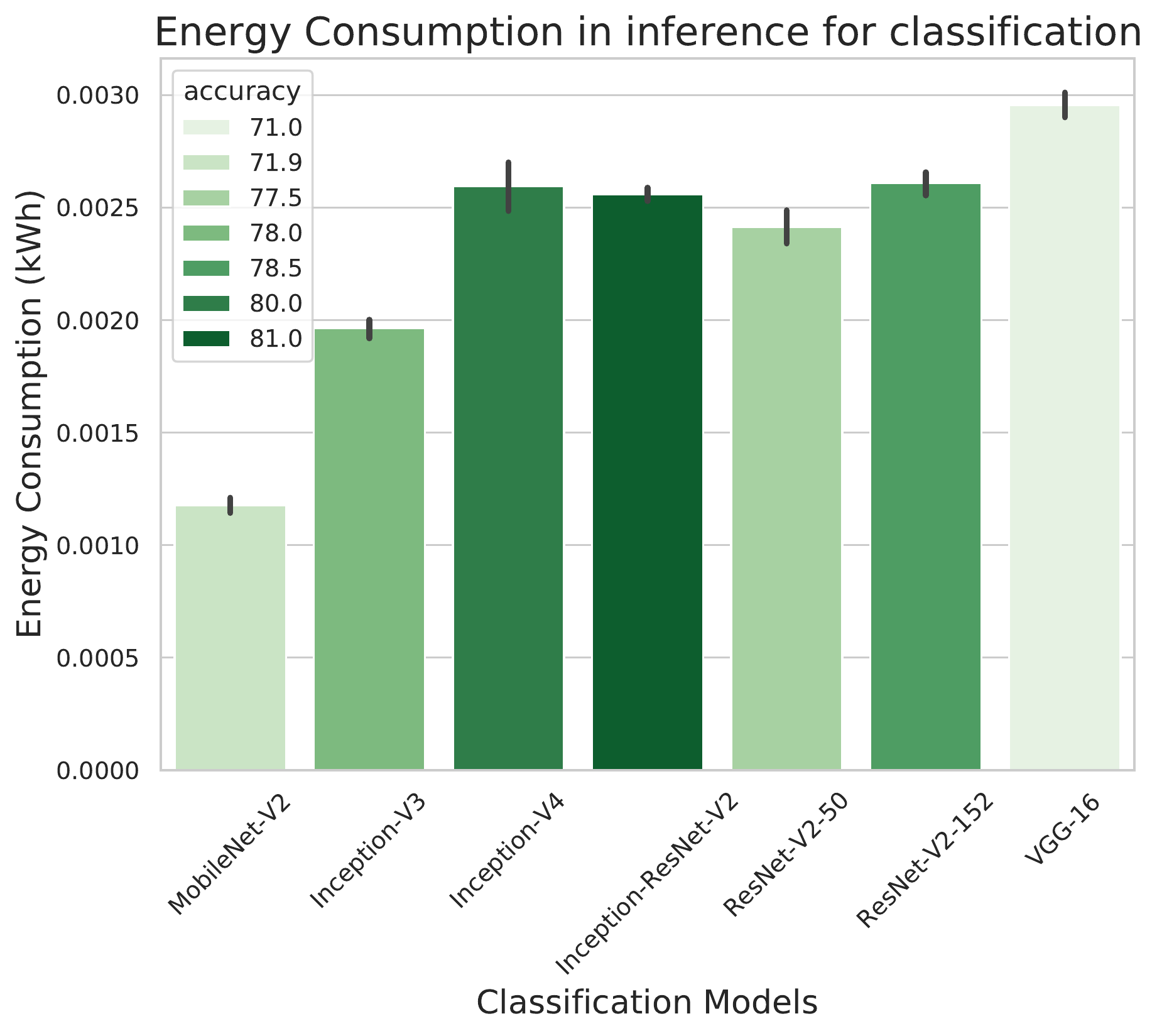}
	\caption{Energy consumption in Inference per image for Classification Models }
	\label{fig:1}
\end{figure}

Figure ~\ref{fig:1} presents the energy consumption in the inference process per image for classification, which shows us the amount of energy it will cost each time we input one more image into the trained relative model.
The small number of parameters and the simple model structure are the main contributors to MobileNet-V2 having the lowest power level. However, this entails that it also has almost the lowest accuracy. The increase in the number of parameters increased the power consumption of Inception-V4 compared to Inception-V3. 

With double the parameters, InceptionResNet-V2 offers 3\% better accuracy than Inception-V3 and 1\%  in Inception-V4. However, it has an energy consumption of approximately the same as Inception-V4 or even lower. It can also be remarked that the relationship between model complexity and energy levels cannot be linear. Because if this happens, the energy consumption of InceptionResNet-V2 will probably double with the corresponding increase in the number of parameters. This result contributes to the hypothesis that the number of parameters of the model does not seem to be the only factor determining the energy consumption. The DNN model's structure also significantly influences the energy it consumes.

VGG-16 had the highest energy consumption among the evaluated models, but it did not offer better accuracy. It is even lower than MobileNet-V2, even though its parameters are approximately 30 times more.

Table~\ref{tab:infe} shows the energy consumption, estimated carbon emission and also the duration time for the inference process. We separately run the inference process on the same dataset with training. Usually, this is not the case when we run on the same dataset. However, it works to simulate the real-life process, and this result still contributes to the conclusion in comparing the energy consumption when we use trained models.
\begin{table}[!h]
\caption{Average Energy Consumption (EC), Carbon Emission (CE), Duration (D) in Inference for Classification Models }
\label{tab:infe}
\centering
\begin{tabular}{l|l|l|l}
  \toprule 
  \textbf{Model} & \textbf{EC (kWh)} & \textbf{CE (kgCO2eq)}& \textbf{D (seconds)}\\
  \midrule 
    MobileNet-V2 & $1,18.10^{-3}$ & $0,66.10^{-4}$&32,11\\
    Inception-V3 & $1,96.10^{-3}$ & $1,10.10^{-4}$&32,26\\
    Inception-V4 & $2,60.10^{-3}$ & $1,45.10^{-4}$&32,41\\
    Inception-ResNet-V2 & $2,56.10^{-3}$ & $1,43.10^{-4}$&32,51\\
    ResNet-V2-50 & $2,42.10^{-3}$ & $1,35.10^{-4}$&32,31\\
    ResNet-V2-152 & $2,61.10^{-3}$ & $1,46.10^{-4}$&32,95\\
    VGG-16 & $2,96.10^{-3}$ & $1,66.10^{-4}$&33,63\\
  \bottomrule 
\end{tabular}
\end{table}

\begin{figure}[!h]
	\centering
	\includegraphics[width=1\linewidth]{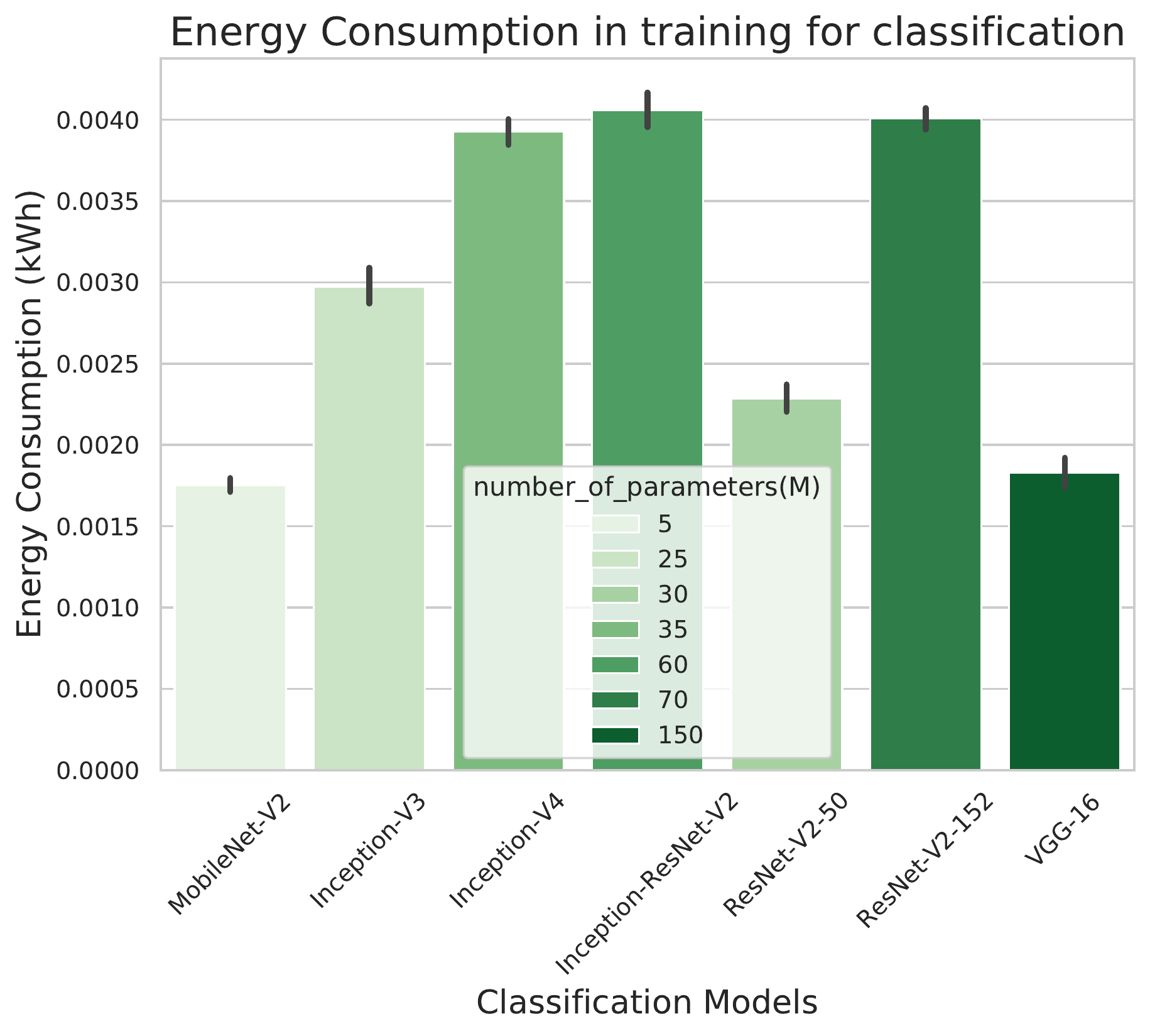}
	\caption{Energy consumption in Training for Classification Models}
	\label{fig:3}
\end{figure}

\begin{figure}[!h]
	\centering
	\includegraphics[width=1\linewidth]{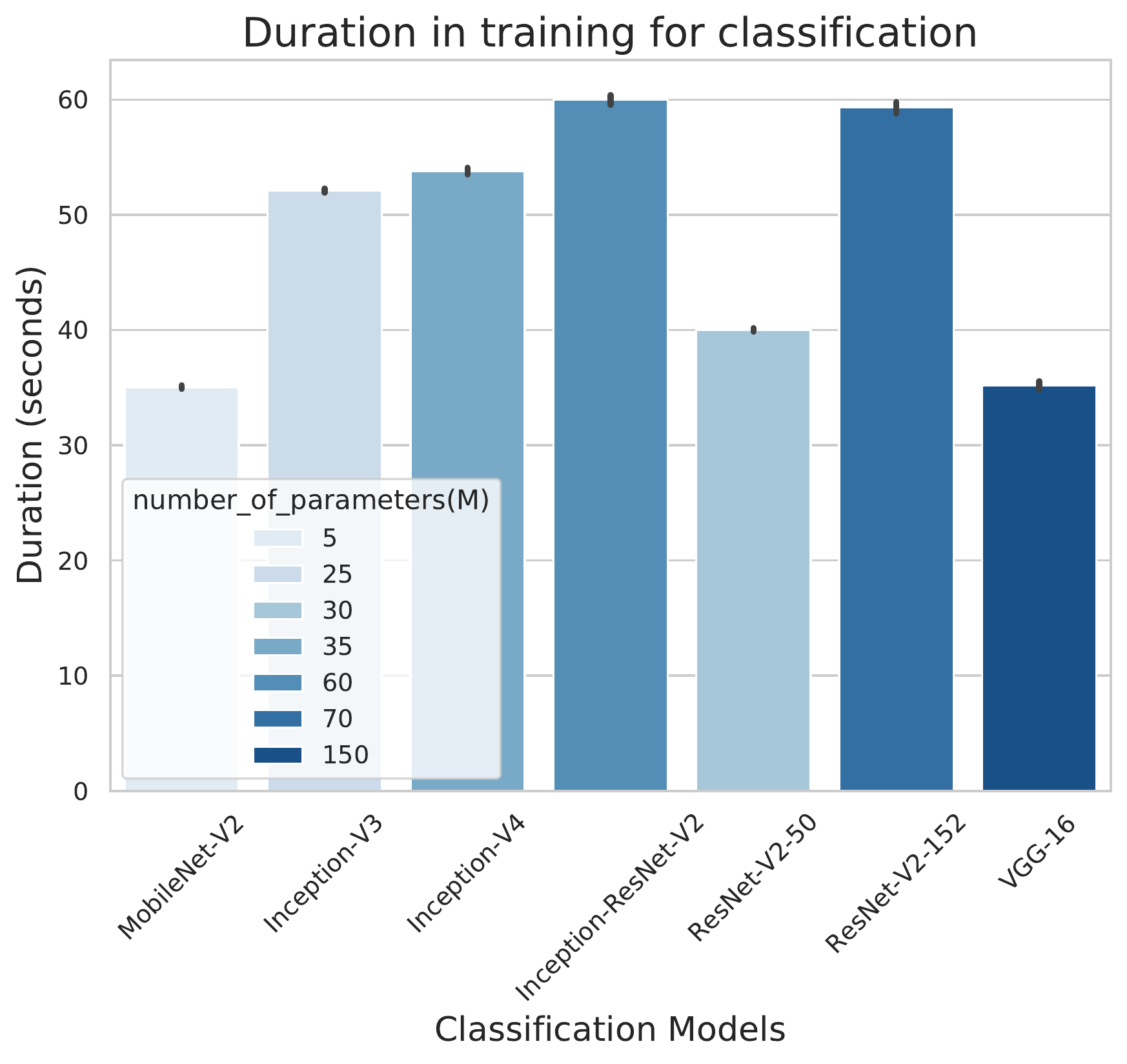}
	\caption{Duration in Training for Classification Models}
	\label{fig:4}
\end{figure}

Comeback with training, Table ~\ref{tab:train} gives the detailed energy consumption scores. From Figure~\ref{fig:3},~\ref{fig:4}, the complexity of the models explains that InceptionResNet-V2, with twice as many parameters for the model as Inception-V3, and Inception-V4,  has the highest energy consumption level. Likewise, the training time for Inception-ResNet-V2 is the biggest among the three models mentioned above. The same goes for ResNet-152, which has twice the number of parameters ResNet-50 and has a higher power and time level than ResNet-50. 

Nevertheless, it did not happen for VGG-16. Even though VGG-16 is the most significant one with 150M parameters, it has remarkably positive results by being the model that consumes the smallest amount of energy and runs in the shortest time. The belonging designed architecture idea of each DNN explains a part of why the energy consumption of VGG-16 and MobileNet-V2  are the best models in terms of energy consumption and training duration for Classification. They both have the same objective when trying to reduce the complexity cost. While MobileNet-V2 reduces the model size in order to run on mobile devices, VGG-16 decreases filter size. This observation shows evidence for the design of deep neural networks can reduce energy consumption and training time.
\begin{table}[!h]
\caption{Average Energy Consumption, Carbon Emission, Duration in Training for Classification Models}
\label{tab:train}
\centering
\begin{tabular}{l|l|l|l}
  \toprule 
  \textbf{Model} & \textbf{EC (kWh)} & \textbf{CE (kgCO2eq)}& \textbf{D (seconds)}\\
  \midrule 
    MobileNet-V2 & $1,75.10^{-3}$ & $0,98.10^{-4}$&35,04\\
    Inception-V3 & $2,98.10^{-3}$ & $1,67.10^{-4}$&52,09\\
    Inception-V4 & $3,93.10^{-3}$ & $2,20.10^{-4}$&53,79\\
    Inception-ResNet-V2 & $4,06.10^{-3}$ & $2,27.10^{-4}$&59,98\\
    ResNet-V2-50 & $2,29.10^{-3}$ & $1,28.10^{-4}$&40,01\\
    ResNet-V2-152 & $4,01.10^{-3}$ & $2,24.10^{-4}$&53,32\\
    VGG-16 & $1,83.10^{-3}$ & $1,02.10^{-4}$&35,19\\
  \bottomrule 
\end{tabular}
\end{table}


\section{Conclusion and future work}
\label{sec:conclusion}
This paper presents a step towards better understanding the energy consumption of AI algorithms, notably Deep Neural Networks (DNNs), when running in High-Performance Computing (HPC) platforms. For this task we instrumented a known AI benchmark with energy measurement tools to create an extended benchmark, called Benchmark-Tracker. 
Benchmark-Tracer works as a new test-bed for evaluating the processing performance and energy consumption of AI algorithms in HPC platforms. For instance, in our case example we found the following observations.

For a certain AI task (in our case example, for the image classification task), more complex DNN models can consume more energy, emit more CO2, and take longer time than simpler models to train. Nevertheless, there are exceptions, and they advise that the energy consumption likewise depends on the structure of the DNNs, and their relative function is not linear. A more accurate model does not necessarily consume more energy. Also, if we bring the connection between inference and training into the balance, the relationship becomes even more complicated. There are cases where the energy level for training is low but exceptionally high in inference and vice versa. 

The choice of model selection and the HPC hardware is not a trivial decision, since it depends on the problem and resources specificities, which are not always well understood theoretically. Benchmark-Tracker can help to perform this decision by performing light-weight experiments to grasp how much energy a certain AI model will consume and how fast it will run, according to a certain HPC hardware.


\subsection{Future work}
For future work we will take into account calculating the accuracy belonging to the models for a specific dataset, besides the energy consumption. This will enable us to perform the following investigations.
\begin{enumerate}
    \item If we reduce or increase the dataset size during training, we can consider how much energy the training will consume and how much accuracy we will get for each model to compare. Furthermore, doing that on a specific model will help understand the tradeoffs of dataset size, energy consumption, and resulting inference accuracy.
    \item With the Benchmark-Tracker, we are going to set energy budgets during the models' training. We can control the training process for a specific objective by stopping the training when the total energy (measured by an energy measurement library) passes a defined budget. Furthermore, as a consequence, we can evaluate the performance behavior of the DNN models when we have limited energy budgets.
    \item Following the energy budget idea, we are going to compare shallow learning algorithms for a selection of applications present in Benchmark-Tracker and whether these shallow learning algorithms outperform or not the deep learning algorithms when we have energy budgets.
\end{enumerate}

\subsubsection{Acknowledgements} 
This work was supported by the research program on Edge Intelligence of the Multi-disciplinary Institute on Artificial Intelligence MIAI at Grenoble Alpes (ANR-19-P3IA-0003). We also thank all institutions (INRIA, CNRS, RENATER and several Universities as well as other organizations) who support the Grid5000 platform.


\bibliographystyle{splncs04}
\bibliography{references}

\end{document}